\documentclass[
aps,
physrev,
preprint,
preprintnumbers,
groupedaddress,
amsmath,
amssymb,
]{revtex4-2}

\usepackage{graphicx}
\usepackage{dcolumn}
\usepackage{bm}
\usepackage[colorlinks=true,
            linkcolor=red,
            citecolor=green,
            urlcolor=gray]{hyperref}
\usepackage[mathlines]{lineno}
\usepackage{subcaption}
\usepackage[justification=justified,singlelinecheck=false]{caption}
\usepackage{ragged2e}
\usepackage{lipsum}
\usepackage{dcolumn}
\usepackage[usenames]{color}
\usepackage{xcolor}
\usepackage{comment}
\usepackage{float}

\begin{document}

\title{\textbf{Loop-Induced Higgs Boson Decays into Gauge Bosons in Radiative Natural Supersymmetry}}
\author{Edilson A. Reyes R.}
\email{Contact author: edilson.reyes@unipamplona.edu.co}
\affiliation{Physics Department - G.O.M, Universidad de Pamplona, Pamplona, Norte de Santander, Colombia. }
\date{\today} 

\begin{abstract}
In this article, we study loop-induced Higgs decays into gauge bosons within the framework of Radiative Natural Supersymmetry. We reproduce the one-loop MSSM calculations for the Higgs partial decay widths into a Z boson-photon pair, two photons, and two gluons, providing the corresponding analytical expressions for the scattering amplitudes. We focus on the region of parameter space that maximizes the rare decay width of the process $h \to Z\gamma$, and analyze the correlated predictions for the remaining Higgs decay channels. In the selected region of parameter space, the $h \to Z\gamma$ decay width is enhanced, reaching a maximum value of $\simeq 7.5~\mathrm{keV}$, while remaining compatible with the current combined ATLAS measurement. At the same time, this region satisfies current Higgs constraints from the $h \to \gamma\gamma$ and $h \to gg$ channels. The diphoton mode remains close to the Standard Model expectation, with deviations at the level of $\lesssim 5\%$. The gluon channel exhibits a stronger sensitivity to the considered region of parameter space, leading to a moderate suppression of about $10\%$ in the corresponding partial width.
\end{abstract}


\maketitle
\tableofcontents

\section{\label{sec:Intro} Introduction}

The discovery of the Higgs boson at the Large Hadron Collider (LHC) completed the particle content of the Standard Model (SM) and marked a milestone in our understanding of electroweak symmetry breaking. As the last fundamental particle of the SM to be observed experimentally, the Higgs boson plays a central role in the consistency of the theory at the quantum level and provides a mechanism for the generation of elementary particle masses. For this reason, a precise determination of its properties, such as its mass, couplings, and decay widths, constitutes a primary goal of current and future collider experiments. \\ \\ From the theoretical perspective, the properties of the Higgs boson have been computed with remarkable precision within the SM and its extensions. In particular, the Higgs boson mass ($M_h$) has been calculated in the Feynman diagrammatic approach reaching multi-loop precision. In the SM, state-of-the-art calculations incorporating higher-order corrections up to three-loop accuracy, have significantly reduced the theoretical uncertainty~\cite{Martin2019,Martin2021,Martin2022,Martin2023,2023HiggsSM}, allowing for a meaningful comparison with experimental measurements where the uncertainty is approximately $100$~MeV~\cite{ATLASCMS,ATLAS2023,CMS2025}. In supersymmetric (SUSY) extensions of the SM, such as the Minimal Supersymmetric Standard Model (MSSM), the Higgs mass prediction is highly sensitive to radiative effects, implying theoretical uncertainties of about one order of magnitude larger than the experimental precision achieved at the LHC ($1-5$~GeV), making precision calculations essential. Advanced multi-loop computations of the Higgs boson mass in SUSY models have been presented in~\cite{2022Particles,2021EPJC} and references therein, where the impact of higher-order corrections and the associated theoretical uncertainties were analyzed in detail. \\ \\ Beyond the Higgs mass, the decay widths of the Higgs boson provide an equally important window into its underlying dynamics. Higgs decay rates are directly sensitive to its couplings and are therefore powerful probes of new physics effects. In the SM, both tree-level and loop-induced Higgs decay channels have been computed with high precision~\cite{Djouadi2007, Spira2017}. Extensions of the SM, including SUSY models, can modify these decay widths through new particles entering radiative corrections or through altered Higgs couplings~\cite{Spira2017, Djouadi2008, CarenaHaber2003}. Experimentally, several Higgs decay channels into gauge bosons have already been measured at the LHC with increasing precision~\cite{HiggsExp1, HiggsExp2}. The three body decays $h \to WW^*\to Wf\bar{f}$ and $h \to ZZ^* \to Zf\bar{f}$ are among the best-measured channels and play a crucial role in Higgs coupling determinations, while $h \to \gamma\gamma$ has provided one of the cleanest discovery signatures. The decay $h \to Z\gamma$, although experimentally challenging due to its small branching ratio~\cite{hgZcomb,hgZAtlas}, remains an important target for future analyses, as its loop-induced nature makes it particularly sensitive to physics beyond the Standard Model (BSM). Future collider projects such as the ILC~\cite{ILC} and the FCC~\cite{FCC} are expected to greatly improve the precision of Higgs measurements, enabling percent-level determinations of Higgs couplings and decay widths. This level of accuracy will substantially enhance the sensitivity to new physics, highlighting the need for precise computations of Higgs properties within BSM scenarios. Any model capable of explaining potential deviations must remain consistent with the well-established SM predictions. \\ \\ In this context, models of radiative natural supersymmetry (RNS)~\cite{Baer2012,Baer2013,Baer2022} provide a well-motivated framework for studying new physics effects in Higgs observables. RNS can be realized within the MSSM without introducing additional exotic matter, accommodating the observed Higgs boson mass while avoiding a further increase in the already sizable theoretical uncertainties associated with it. Moreover, RNS preserves electroweak naturalness and is consistent with current LHC constraints from $B$ physics and SUSY particle searches. Particularly, a detailed study of the $h \to Z\gamma$ decay was performed within the framework of RNS in reference~\cite{Edilson2025}. In that work, the corresponding Higgs decay width ($\Gamma_{Z\gamma}$) was analyzed using a region of parameter space where $\Gamma_{Z\gamma}$ is maximized while respecting important experimental constraints. The $\mu$ parameter was constrained to values above $100~\text{GeV}$, in accordance with the current ATLAS and CMS lower limits on chargino masses. Within this allowed range, small values of $\mu$ are favored by electroweak naturalness, since they lead to a reduced fine-tuning parameter. Furthermore, low fine-tuning scenarios in RNS also maintain consistency with flavor observables such as the $b\to s\gamma$ branching ratio. Besides, it was shown that SUSY contributions, arising mainly from chargino loops, can lead to deviations of about $20\%$ with respect to the SM prediction ($\Gamma_{Z\gamma}^{\text{SM}} \sim 6.2~\text{keV}$). This is consistent with the latest combined Run 2 + Run 3 ATLAS result~\cite{ATLAS2025}, $\Gamma_{Z\gamma}^{\mathrm{ATLAS}}\sim 8.1^{+3.6}_{-3.2}~\mathrm{keV}$, as well as with the latest CMS result~\cite{CMS2026}, $\Gamma_{Z\gamma}^{\mathrm{CMS}}\sim 6.8^{+3.2}_{-3.8}~\mathrm{keV}$. Both measurements still exhibit large uncertainties, leaving room for potential new physics effects in this channel. At the same time, we emphasize that any modification of the $h\to Z\gamma$ rate must be confronted with other precision observables. In particular, electroweak precision tests, such as the oblique parameters $S,T,U$ and $m_{W}$, as well as $Z$-pole observables and the other Higgs signal strengths could severely constrain new-physics effects. Having this in mind, in the present work we extend the analysis in~\cite{Edilson2025} to a comprehensive study of the loop-induced Higgs boson decays into pairs of gauge bosons, $h \to V_1 V_2$, within the same RNS framework. With this analysis we determine whether the enhancement in $h\to Z\gamma$ can be realized without spoiling the successful SM-like predictions in other channels, and we obtain a coherent picture of the impact of RNS on Higgs phenomenology. \\ \\ This paper is organized as follows. In section~\ref{sec:RNS} we briefly review the theoretical framework of RNS and summarize the parameter space considered in our analysis. In section~\ref{sec:decays} we present the formalism for Higgs decays into a pair of gauge bosons and describe the calculations of the corresponding loop-induced decay widths at one-loop level. Numerical results are discussed in section~\ref{sec:numerical}. Finally, section~\ref{sec:conclusions} contains our conclusions. The main routines and data that produce the results presented in this paper are available at~\cite{data}. 

\section{\label{sec:RNS} Parameter Scan in RNS}

Given that Higgs boson decay widths are sensitive to the masses and couplings of both SM and BSM particles, it is important to study how these observables vary as functions of the relevant new-physics scales, and to quantify the size of the resulting deviations with respect to the SM predictions. In this paper, we focus on the effects of MSSM corrections, imposing that the numerical values of the SUSY mass spectrum and the relevant couplings entering the predictions of Higgs decay widths are obtained through their renormalization-group evolution. The running is controlled by the renormalization-group equations (RGEs) with boundary conditions defined in terms of a constrained set of unified input parameters specified at the grand unification scale, $\Lambda_{\mathrm{GUT}} = 1.5 \times 10^{16}\,\text{GeV}$, within the RNS framework. These RNS inputs include the universal scalar mass $m_0$, the universal gaugino mass $m_{1/2}$, the universal trilinear coupling $A_0$, the ratio of the vacuum expectation values of the two Higgs doublets in the MSSM, $\tan\beta = v_u/v_d$, the Higgsino mass parameter $\mu$, and the mass of the CP-odd Higgs boson $m_A$. We consider the region of parameters that maximizes the value of the decay width for the process $h\to Z\gamma$, while preserving the Higgs mass prediction in MSSM inside the range $M_h = 125 \pm 3 ~ \text{GeV}$ and a moderately large fine-tuning parameter $\Delta_{EW} \approx 100$, as was discussed in reference~\cite{Edilson2025}. Thus, we scan the following parameter space:
\begin{eqnarray}
50\:\text{GeV}\leq\mu\leq350\:\text{GeV}, & \quad 600\,\text{GeV}\leq m_{A}\leq1500\:\text{GeV}, & \quad 500\:\text{GeV}\leq m_{1/2}\leq2500\:\text{GeV}, \nonumber \\
 \left|A_{0}/m_{0}\right|=1.75, \qquad & 2\:\text{TeV}\le m_{0}\leq8\:\text{TeV} , & \qquad 3.5\:\text{TeV}\leq \left|A_{0}\right| \leq 14\:\text{TeV},  \nonumber \\
 & \quad 8\leq\tan\beta \leq55\, \label{eq:RNSscan} . & 
\end{eqnarray}

\noindent We use the SUSY spectrum generator \texttt{ISASUGRA}~\cite{Isasugra}, included in the \texttt{ISAJET} package~\cite{Isajet}, to perform the evolution of the MSSM parameters. The \texttt{ISAJET} code integrates the RGEs of the MSSM parameters, including the full two-loop contributions for gauge and Yukawa couplings, together with a consistent treatment of the soft SUSY-breaking parameters. Within the \texttt{ISAJET} convention, a commonly adopted choice for the SUSY matching scale is the geometric mean of the stop masses, $Q_{\rm SUSY} = \sqrt{m_{\tilde t_1}\,m_{\tilde t_2}}$. The MSSM parameters are evolved from $\Lambda_{\rm GUT}$ down to $Q_{\rm SUSY}$, at which the soft-breaking parameters and running couplings are determined. This scale is chosen to minimize large logarithmic corrections induced by the stop sector. The resulting parameters are then used to construct the physical SUSY spectrum. Internally, the running parameters follow the standard $\overline{\mathrm{DR}}/\text{DRED}$ renormalization scheme~\cite{DRED,CAPPER,Dominik2005} and are consistently converted into physical pole masses by including the relevant loop corrections. 
\begin{table*}[t]
\centering
\begin{tabular}{|c|c|c|c|}
\hline
\textbf{Parameter} & \textbf{BP1} & \textbf{BP2} & \textbf{BP3} \\ \hline
\multicolumn{4}{|c|}{\texttt{Input parameters at $\Lambda_{\rm GUT}$}}\\ \hline
$m_0$ (TeV)             & 2.0   & 4.0   & 8.0   \\ 
$A_0$ (TeV)             & -3.5  & -7.0  & -14.0 \\ \hline
\multicolumn{4}{|c|}{\texttt{Low-energy parameters at $Q_{\rm SUSY}$}}\\ \hline
$A_t$ (GeV)                         & $-3659.8$ & $-4942.6$ & $-7479.0$ \\
$A_{\tau}$ (GeV)                    & $-3600.3$ & $-6506.1$ & $-12357$ \\
$m_{\tilde g}$ (GeV)                & $3265.8$  & $3380.6$  & $3553.7$ \\
$m_{\tilde\chi_1^0}$ (GeV)          & $98.908$   & $99.116$   & $99.564$ \\
$m_{\tilde\chi_1^\pm}$ (GeV)        & $106.11$  & $106.42$  & $107.27$ \\
$m_{\tilde\chi_2^\pm}$ (GeV)        & $1227.2$ & $1254.6$ & $1293.8$ \\
$m_{\tilde t_1}$ (GeV)              & $1739.7$ & $1610.6$ & $2108.5$ \\
$m_{\tilde t_2}$ (GeV)              & $2663.6$ & $3305.1$ & $5339.9$ \\
$m_{\tilde b_1}$ (GeV)              & $2651.4$ & $3316.2$ & $5382.0$ \\
$m_{\tilde\tau_1}$ (GeV)            & $1589.6$ & $3259.2$ & $6616.3$ \\ \hline
$M_h$ (GeV)                         &    123.78   &  125.69    &  127.72   \\
$\Delta_{\rm EW}$                   &   50.970    &  48.980    &  204.56     \\ \hline
\end{tabular}
\caption{ \justifying \small{Representative benchmark points and the corresponding low-energy spectrum used to evaluate the loop-induced $h\to V_1V_2$ decay amplitudes. All benchmark points reproduce the Higgs boson mass in the range $M_h=125\pm3~\mathrm{GeV}$, and satisfy the current collider constraints.}}
\label{tab:BP}
\end{table*} \\ 
\noindent Table~\ref{tab:BP} presents three representative benchmark points, listing the GUT-scale input parameters together with the relevant low-energy SUSY spectrum entering the dominant Higgs decay amplitudes. The benchmarks are defined by varying the universal scalar mass $m_0$ from $2$ to $8~\text{TeV}$, while maintaining the common RNS input parameters $\mu=100~{\rm GeV}$, $m_{1/2}=1.5~{\rm TeV}$, $m_A=600~{\rm GeV}$, and $\tan\beta=30$. The trilinear coupling $A_0$ is fixed such that $|A_0/m_0|=1.75$. With this choice, the lightest chargino and neutralino masses remain nearly constant over the three benchmark points, with $m_{\tilde{\chi}_1^\pm}\simeq106$--$107~{\rm GeV}$ and $m_{\tilde{\chi}_1^0}\simeq99~{\rm GeV}$ respectively. Similarly, the gluino mass $m_{\tilde g}$ remains above $3.2~{\rm TeV}$ for all benchmark points, while the scalar sector exhibits a much stronger dependence on $m_0$, particularly in the stop and stau masses. The lightest stop mass $m_{\tilde t_1}$ varies from about $1.6$ to $2.1~{\rm TeV}$, whereas the heavier stop ($\tilde{t}_2$) and the lightest sbottom ($\tilde{b}_1$) and stau ($\tilde{\tau}_1$) masses increase substantially as $m_0$ is raised. As a consequence, the loop-induced Higgs decay widths are expected to be affected predominantly by the scalar sector in this region of parameter space. Note that all benchmark points satisfy the Higgs boson mass requirement, while the electroweak fine-tuning remains of order $\mathcal{O}(100)$, and are consistent with current collider constraints~\cite{PDG2026}. In particular, the gluino and third-generation sfermion masses lie well above the present exclusion limits from ATLAS and CMS searches, while the nearly degenerate higgsino-like chargino and neutralino spectrum remains consistent with current LHC searches for electroweak supersymmetric scenarios. \\ 
In the context of RNS, the requirement of natural electroweak symmetry breaking implies that $\mu$ must lie close to the electroweak scale, while experimental constraints from LEP2 and the LHC exclude charginos with masses below approximately $103.5\,\mathrm{GeV}$. Consequently, RNS scenarios demand $\mu \gtrsim 100\,\mathrm{GeV}$, leading to a lightest neutralino (the MSSM predicts a spectrum where the lightest SUSY particle is typically the lightest neutralino $\widetilde{\chi}_1^0$) that is almost purely Higgsino-like, assuming the limit where gaugino mass parameters $M_1$ (bino) and $M_2$ (wino) are much heavier than the Higgsino mass $\mu$. All other superpartners, including heavier neutralinos $\tilde{\chi}_{j=2,3,4}^0$, charginos $\tilde{\chi}_{i=2}^\pm$, sleptons $\tilde{l}_j$ and squarks $\tilde{q}_j$, have significantly larger masses. This implies that decays of the Higgs boson into SUSY fermions, such as $h \to \tilde{\chi}_i^0 \tilde{\chi}_j^0$ and $h \to \tilde{\chi}_i^\pm \tilde{\chi}_j^\mp$, as well as into SUSY scalars, such as $h \to \tilde{l}_i \tilde{l}_j$ and $h \to \tilde{q}_i \tilde{q}_j$, are kinematically forbidden. In particular $m_{\widetilde{\chi}_1^0} > M_h / 2$ preventing the decay $h \rightarrow \widetilde{\chi}_1^0 \widetilde{\chi}_1^0$, and all other sparticles are well above this threshold. First- and second-generation sfermions are generally decoupled at multi-TeV scales to comply with LHC null search results, while third-generation squarks, though possibly lighter, are still too heavy to allow for kinematically viable decays of the 125~GeV Higgs boson. As a consequence, by imposing the RNS boundary conditions the Higgs boson decays only into SM particles and the total width of the Higgs boson ($\Gamma_h$) receives no direct contribution from decays into SUSY particles. This makes the precise computation of the radiative corrections of these Higgs partial widths particularly relevant, as any deviations from the SM predictions must arise mainly from loop-level effects, especially in the decoupling limit~\cite{Djouadi2008, Dobado} satisfied in RNS ($m_A \gg M_Z$ and $\alpha \approx \beta - \frac{\pi}{2}$) where the couplings of the light CP-even Higgs boson ($h$) to SM fermions and gauge bosons approach their SM values, up to corrections of order $M_Z^2/m_A^2$. As a result, detailed studies of the loop-induced processes $h \to \gamma\gamma$, $h \to Z\gamma$, and $h \to gg$ provide particularly sensitive probes of SUSY effects, as they directly probe the presence of superpartners in the loops. In addition, analyzing their partial decay widths offers complementary information beyond a purely $\kappa$-parameter-based approach, which is commonly used in global fits. A key feature of partial decay widths is the presence of interference effects among different loop contributions in the decay amplitude. Strong destructive interference between the $W$-boson and top-quark contributions has been reported for the $h \to \gamma\gamma$~\cite{Ellis1976,Shifman2012} and $h \to Z\gamma$~\cite{Cahn1979,Bergstrom1985,Spira1992,Gehrmann2015} decays at one-loop level. Similar cancellations occur at higher orders between the QCD and electroweak contributions~\cite{Chen2024,Sang2024,Sang2025}. These interference patterns imply that even small additional contributions from new physics can lead to sizable effects on the decay widths.

\section{\label{sec:decays} Higgs Decays into Gauge Bosons}

In this section we consider the decays of the Higgs boson into pairs of gauge bosons within the MSSM. We summarize the formalism and computational setup used to evaluate the corresponding decay amplitudes and partial widths, as well as the assumptions underlying the RNS parameter space adopted in our analysis. Main results for the individual decay channels are also discussed. \\ \\ The decay width for the Higgs boson $h$ decaying into a pair of gauge bosons $V_1$ and $V_2$ is obtained from the squared scattering matrix element $\mathcal{M}$ using the standard two-body phase-space master formula,
\begin{equation}
\Gamma(h \to V_1 V_2)
= \frac{1}{16\pi M_h}
\lambda^{1/2}\!\left(1,\frac{M_{V_1}^2}{M_h^2},\frac{M_{V_2}^2}{M_h^2}\right)
\overline{|\mathcal{M}_{V_1 V_2}|^2},
\end{equation}
where $\lambda(x,y,z)=x^2+y^2+z^2-2(xy+xz+yz)$ is the Källen function and
\begin{equation}
\mathcal{M}_{V_1 V_2} = \mathcal{M}_{V_1 V_2}^{\text{SM-like}} + \mathcal{M}_{V_1 V_2}^{\text{SUSY}} \, .
\end{equation}
We define $\mathcal{M}_{ij}^{\text{SM-like}}$ to include all contributions from SM-like particles within the MSSM, while $\mathcal{M}_{ij}^{\text{SUSY}}$ contains only amplitudes involving superpartners. In the decoupling limit under consideration, $\mathcal{M}_{ij}^{\text{SM-like}}$ coincides with the SM amplitude. The corresponding amplitudes have been extensively studied in the literature. For completeness, in this work we reproduce the calculations of the decay widths $\Gamma(h \to V_1 V_2)\equiv \Gamma_{V_1V_2}$ in the $R_{\zeta}$ gauge at the one-loop level. The Feynman diagrams and the corresponding amplitudes contributing to $\mathcal{M}_{V_1 V_2}$ are generated in four dimensions using the \texttt{Mathematica} package \texttt{FeynArts}~\cite{FeynArts}. The leading-order contributions to the loop-induced processes $h\to gg$, $h\to \gamma\gamma$, and $h\to Z\gamma$ arise at one-loop order, and each corresponding amplitude $\mathcal{M}$ is ultraviolet finite and does not require additional vertex renormalization. Although one-loop diagrams for $h \to g\gamma $ and $h \to gZ$ can be formally generated with \texttt{FeynArts}, these decays are forbidden at the effective-field-theory level, since gauge invariance severely restricts the possible local operators. In particular, an operator involving a single gluon field strength and the Higgs scalar, such as $hG_{\mu\nu}^{a}F^{\mu\nu}$ or $hG_{\mu\nu}^{a}Z^{\mu\nu}$, cannot be constructed without violating the gauge symmetry and therefore does not induce physical Higgs decays. The resulting amplitudes are regularized in $D$ dimensions using dimensional regularization and manipulated with the \texttt{FeynCalc} package~\cite{FeynCalc}, which is used to perform the required algebraic operations. Through the \texttt{FeynHelpers} interface~\cite{FeynHelpers}, the outputs of \texttt{FeynCalc} are connected to the program \texttt{Package-X}~\cite{PackageX}, allowing the loop integrals to be expressed in terms of Passarino--Veltman functions and enabling the separation of the finite and divergent contributions of each diagram. The numerical evaluation of the involved loop functions can be straightforwardly performed using \texttt{LoopTools}~\cite{LoopTools}.

\subsection{\label{sec:contributions} Dominant Contributions to Higgs Decay Widths}

In the following, we present the form factors for the dominant contributions entering the computation of the scattering amplitudes. By using the Ward-Takahashi identity and the transversality condition of the polarization vectors, the scattering amplitude at one-loop level can be written as
\begin{equation}
\mathcal{M}_{V_1 V_2} \approx
\varepsilon^{\lambda_1}_\mu(V_1)\, \varepsilon^{\lambda_2}_\nu(V_2)\,
\left[
\mathcal{A}_{V_1 V_2}^{(1)}\, g^{\mu\nu}
+ \mathcal{A}_{V_1 V_2}^{(2)}\, \left(k_1 \cdot k_2\, g^{\mu\nu} - k_1^\nu k_2^\mu\right)
\right]. \label{eq:M}
\end{equation}
$\mathcal{A}_{V_1 V_2}^{(j)}$ denotes reduced form factors. The overall process-dependent electroweak and QCD prefactors, including the gauge couplings and color factors, are not included explicitly in equation~(\ref{eq:M}), they are restored in the corresponding decay-width expressions. For all loop-induced processes, the form factor $\mathcal{A}_{V_1 V_2}^{(1)}$ vanishes. Although we do not reproduce here the calculations of Higgs decays into electroweak gauge bosons, it is worth noting that in the three-body decays $h \to VV^* \to V f \bar{f}$, with $V = Z, W$, the form factor $\mathcal{A}_{V_1 V_2}^{(1)}$ provides the dominant contribution, while $\mathcal{A}_{V_1 V_2}^{(2)}$ is generally subdominant. 

\subsubsection{$h \to \gamma\gamma$ decay}

\begin{figure}[]
    \centering
    \includegraphics[width=0.7\textwidth]{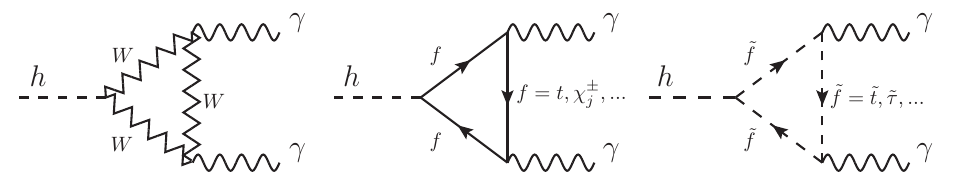}
    \caption{ 
        \justifying \small{Main diagrams contributing to the scattering amplitude $\mathcal{M}_{\gamma \gamma}$ in the $h \to \gamma\gamma$ decay. The internal wavy lines represent $W^{\pm}$ bosons, the solid lines represent the fermionic contributions and the internal dashed lines represent sfermion fields. }
       }
    \label{fig:Hphph}
\end{figure}

The dominant contribution to $\mathcal{M}_{\gamma \gamma}$ arises from the $W$-boson loop, while the top-quark loop provides a sizable destructive interference. In the MSSM, additional contributions from stop and chargino loops can be relevant, with their relative impact determined by the interplay between stop mixing and the presence of light charginos. Light staus can also become important, particularly in regions with large left-right mixing or enhanced Yukawa couplings. The relevant Feynman diagram contributing to the calculation of $\mathcal{M}_{\gamma \gamma}$ are depicted in Figure~\ref{fig:Hphph}.
The decay amplitude can be written in terms of the form factor
\begin{align}
\mathcal{A}_{\gamma\gamma}^{(2)} & =g_{hWW}F_{1}(\tau_{W})+\sum_{f=l,q}N_{c}Q_{f}^{2}\,g_{hff}F_{1/2}(\tau_{f}) \nonumber \\
 & \quad + \sum_{\tilde{\chi}^{\pm}}g_{h\tilde{\chi}^{+}\tilde{\chi}^{-}}F_{1/2}(\tau_{\tilde{\chi}^{\pm}})+\sum_{\tilde{f}=\tilde{l},\tilde{q}}N_{c}Q_{\tilde{f}}^{2}\,g_{h\tilde{f}\tilde{f}}F_{0}(\tau_{\tilde{f}}),
\end{align}
where $\tau_i = 4m_i^2/M_h^2$. The factor $N_c$ denotes the color multiplicity of the particle in the loop. For colored particles such as quarks ($q$) and squarks ($\tilde{q}$), $N_c = 3$, while for colorless particles such as leptons ($l$), sleptons ($\tilde{l}$), and charginos ($\chi^{\pm}$), $N_c = 1$. The quantity $Q_i$ represents the electric charge of the particle $i$ in units of the proton charge. The parameters $g_{hXY}$ denote the effective Higgs couplings to the particles running in the loop and can be consulted in Appendix~\ref{sec:ap_A}. The functions entering the form factor $\mathcal{A}_{\gamma\gamma}^{(2)}$ are defined as
\begin{align}
F_0(\tau) &= -\tau \left[1 - \tau f(\tau)\right], \label{F_0} \\
F_{1/2}(\tau) &= 2\tau \left[1 + (1-\tau) f(\tau)\right], \label{F_1/2}\\
F_1(\tau) &= -\left[2 + 3\tau + 3\tau(2-\tau) f(\tau)\right]. \label{F_1}
\end{align}
The function $f(\tau)$ is given by
\begin{equation}
f(\tau) =
\begin{cases}
\arcsin^2\left(\frac{1}{\sqrt{\tau}}\right), & \tau \geq 1, \\
-\frac{1}{4} \left[ \ln\left(\frac{1+\sqrt{1-\tau}}{1-\sqrt{1-\tau}}\right) - i\pi \right]^2, & \tau < 1.
\end{cases}
\end{equation}
The $h \to \gamma\gamma$ partial decay width is obtained from
\begin{equation}
    \Gamma_{\rm \gamma\gamma} = \frac{G_F \alpha^2 M_h^3 }{128 \sqrt{2}\pi^3} \left| \mathcal{A}_{\rm \gamma\gamma}^{(2)} \right|^2 . 
\end{equation} 
In the RNS regime, stop contributions to $\mathcal{M}_{\gamma\gamma}$ are typically suppressed due to their relatively large masses, while stau contributions are generally subleading unless very light sleptons are present. Chargino contributions can become relevant due to the light higgsino states. However, their impact is typically moderate, as the corresponding couplings are controlled by the higgsino-gaugino mixing, which is limited in scenarios with heavy gauginos. As a result, the overall amplitude remains largely dominated by the $W$ loop, with small corrections from fermionic contributions.

\subsubsection{$h \to Z\gamma$ decay}
\begin{figure}[]
    \centering
    \includegraphics[width=0.7\textwidth]{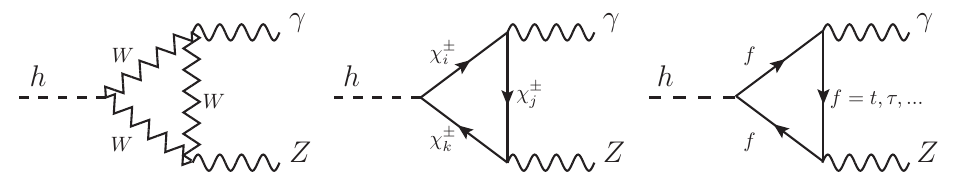}
    \caption{ 
        \justifying \small{Feynman diagrams contributing to the dominant part of the scattering amplitude $\mathcal{M}_{Z\gamma}$ for the decay $h \to Z\gamma$.}
       }
    \label{fig:HphZ}
\end{figure}

The dominant contributions in the MSSM to the scattering amplitude $\mathcal{M}_{Z \gamma}$ comes from the Feynman diagrams drawn in Figure~\ref{fig:HphZ}.
As in the diphoton channel, the $W$ loop dominates, while the top-quark contribution is suppressed but still induces a non-negligible destructive interference. This decay is particularly sensitive to the electroweak particles running in the loop, making chargino contributions especially relevant in RNS. The dominant form factor takes the form
\begin{align}
\mathcal{A}_{Z\gamma}^{(2)} & \simeq g_{hWW}g_{ZWW}F_{1}(\tau_{W},\lambda_{W})+\sum_{f}g_{hff}g_{Zff}F_{1/2}(\tau_{f},\lambda_{f}) \nonumber\\
 & \quad+\sum_{\tilde{\chi}^{\pm}}g_{h\tilde{\chi}^{+}\tilde{\chi}^{-}}g_{Z\tilde{\chi}^{+}\tilde{\chi}^{-}}F_{1/2}(\tau_{\tilde{\chi}^{\pm}},\lambda_{\tilde{\chi}^{\pm}})+\sum_{\tilde{f}}g_{h\tilde{f}\tilde{f}}g_{Z\tilde{f}\tilde{f}}F_{0}(\tau_{\tilde{f}},\lambda_{\tilde{f}}),
\end{align}
with $\tau_i = 4m_i^2/M_h^2$ and $\lambda_i = 4m_i^2/M_Z^2$. For the $h \to Z\gamma$ decay, the loop functions can be expressed in terms of the auxiliary functions $I_1$ and $I_2$:
\begin{align}
F_{1/2}(\tau,\lambda) &= I_1(\tau,\lambda) - I_2(\tau,\lambda), \\
F_0(\tau,\lambda) &= I_1(\tau,\lambda),
\end{align}
\begin{align}
F_1(\tau,\lambda) &= c_W \Bigg\{ 4\left(3-\tan^2\theta_W\right) I_2(\tau,\lambda) \nonumber \\
&\quad + \left[\left(1+\frac{2}{\tau}\right)\tan^2\theta_W - \left(5+\frac{2}{\tau}\right)\right] I_1(\tau,\lambda) \Bigg\},
\end{align}
where
\begin{align}
I_1(\tau,\lambda) &= \frac{\tau\lambda}{2(\tau-\lambda)} 
+ \frac{\tau^2 \lambda^2}{2(\tau-\lambda)^2}\left[f(\tau)-f(\lambda)\right] \nonumber \\
&\quad + \frac{\tau^2 \lambda}{(\tau-\lambda)^2}\left[g(\tau)-g(\lambda)\right],
\end{align}
\begin{equation}
I_2(\tau,\lambda) = -\frac{\tau\lambda}{2(\tau-\lambda)}\left[f(\tau)-f(\lambda)\right].
\end{equation}
The function $g(\tau)$ is defined as
\begin{equation}
g(\tau) =
\begin{cases}
\sqrt{\tau-1}\,\arcsin\left(\frac{1}{\sqrt{\tau}}\right), & \tau \geq 1, \\
\frac{\sqrt{1-\tau}}{2} \left[ \ln\left(\frac{1+\sqrt{1-\tau}}{1-\sqrt{1-\tau}}\right) - i\pi \right], & \tau < 1.
\end{cases}
\end{equation}
The $h \to Z\gamma$ partial decay width is obtained from
\begin{equation}
    \Gamma_{Z\gamma} = \frac{G_F^2 M_W^2 \alpha M_h^3 }{64 \pi^4} \left( 1 - \frac{M_Z^2}{M_h^2} \right)^3 \left| \mathcal{A}_{\rm Z\gamma}^{(2)} \right|^2 . 
\end{equation}
In RNS scenarios, the presence of light higgsinos can enhance the chargino contribution to the point where it exceeds that of the top quark, while still remaining subdominant with respect to the $W$-boson loop. Consequently, the chargino loop typically provides the leading supersymmetric correction to the amplitude in this channel.

\subsubsection{$h \to gg$ decay}

\begin{figure}[]
    \centering
    \includegraphics[width=0.6\textwidth]{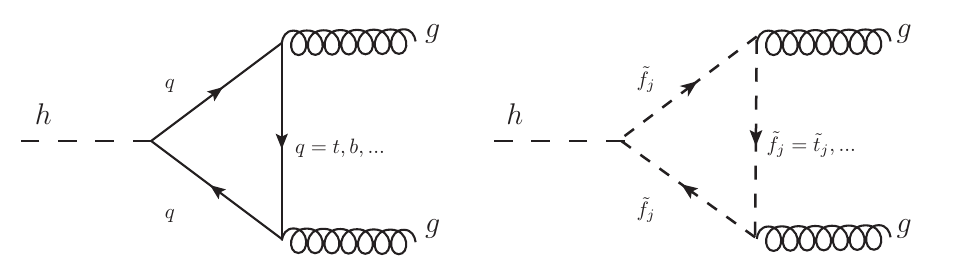}
    \caption{ 
        \justifying \small{Representative Feynman diagrams for the dominant contributions to the amplitude $\mathcal{M}_{gg}$ in the $h \to gg$ decay. Solid (dashed) internal lines denote fermions (sfermions). }
       }
    \label{fig:Hglgl}
\end{figure}

The dominant contribution to $\mathcal{M}_{gg}$, shown in Figure~\ref{fig:Hglgl}, arises from the top-quark loop, with a smaller destructive contribution from the bottom quark.
In the MSSM, scalar contributions can significantly modify the amplitude, especially in scenarios with light sfermions and large mixing. The corresponding form factor is given by
\begin{equation}
\mathcal{A}_{\rm gg}^{(2)} =
\sum_q g_{hqq} F_{1/2}(\tau_q)
+ \sum_{\tilde{q}} g_{h\tilde{q}\tilde{q}} F_0(\tau_{\tilde{q}}),
\end{equation}
where functions $F_{0}$ and $F_{1/2}$ are defined as in (\ref{F_0}) and (\ref{F_1/2}) respectively. The $h \to gg$ partial decay width is obtained from
\begin{equation}
    \Gamma_{\rm gg} = \frac{G_F \alpha_s^2 M_h^3 }{36 \sqrt{2}\pi^3} \left| \mathcal{A}_{\rm gg}^{(2)} \right|^2 . 
\end{equation}
In the RNS framework, stop contributions can modify the decay width if one of the stop eigenstates is relatively light. Depending on the stop masses and mixing parameters, the scalar contribution may interfere constructively or destructively with the dominant top-quark loop, leading to either enhancements or suppressions of $\mathcal{A}_{\rm gg}^{(2)}$. However, for typical RNS spectra, these effects remain subleading compared to the dominant top-quark contribution. It is worth mentioning that the branching ratio of $h \to gg$ is not directly measured as a final-state decay mode at the LHC, the same effective coupling governs the dominant Higgs production mechanism via gluon fusion. Therefore, precise predictions for $\Gamma_{\rm gg}$ provide an indirect probe of Higgs production dynamics.

\section{\label{sec:numerical} Numerical Analysis \protect}

\begin{figure}[]
    \centering
    \includegraphics[width=0.55\textwidth]{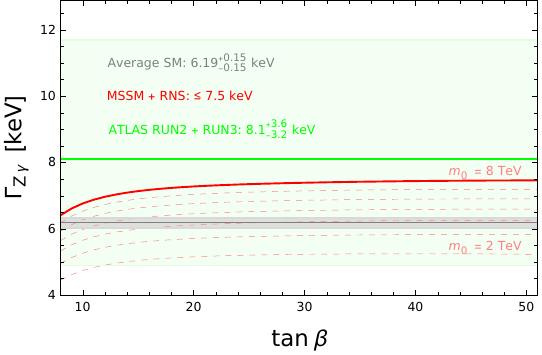}
    \caption{ 
        \justifying \small{$h \to Z\gamma$ decay width as a function of $\tan\beta$ for different values of $m_0$ in the RNS scenario. The gray band represents the SM prediction with its theoretical uncertainty, while the green band corresponds to the combined RUN2 + RUN3 ATLAS measurement.}
       }
    \label{fig:GhZg}
\end{figure}

We begin our numerical analysis by focusing on the region of the RNS parameter space that maximizes the $h \to Z\gamma$ decay width. The predictions for $\Gamma_{Z\gamma}$ in this region are shown in Figure~\ref{fig:GhZg}, where solid and dashed red curves represent the values of $\Gamma_{Z\gamma}$ for different choices of $m_0$ and $\tan\beta$. In particular, the solid red line corresponds to the RNS configuration that maximizes the prediction for $\Gamma_{Z\gamma}$ with $m_0 = 8~\mathrm{TeV}$, $A_0 = -14~\mathrm{TeV}$, $\mu=100~\text{GeV}$ and $m_{1/2}=1.5~\text{TeV}$. For $\tan\beta = 51$, the prediction reaches its maximum value of $\Gamma_{Z\gamma} = 7.5~\mathrm{keV}$.   
We include in this plot the average SM prediction and its theoretical uncertainty, estimated in Ref.~\cite{Edilson2025} as $\Gamma_{Z\gamma}^{\rm SM} = 6.19 \pm 0.15~\mathrm{keV}$. The central value is shown as a gray line, while the associated uncertainty is represented by a gray band. We also include the latest combined RUN2 + RUN3 ATLAS result, depicted as a green line. The ATLAS measurement exhibits relative uncertainties of approximately $40\%$--$45\%$, represented by a green band. Due to the large experimental uncertainty, the current ATLAS measurement is compatible with both the SM and RNS predictions, preventing a clear discrimination between the two scenarios. A reduction in the experimental uncertainty would be essential to enhance the sensitivity to new physics effects and to test whether the RNS scenario can be distinguished from the SM expectation. To be more precise, taking the central value of the measurement and assuming that it remains stable as the experimental precision improves, the difference with respect to the SM prediction is $\Delta \Gamma \simeq 1.9~\mathrm{keV}$. In order for the SM prediction to lie outside the $1\sigma$ experimental interval, the uncertainty would need to be reduced to $\sigma_{\mathrm{exp}} \lesssim 1.9~\mathrm{keV}$, corresponding to a relative uncertainty of approximately $24\%$. A more stringent discrimination at the $2\sigma$ level requires $\sigma_{\mathrm{exp}} \lesssim 1.0~\mathrm{keV}$, i.e., a relative uncertainty of about $12\%$, while a $3\sigma$ exclusion would demand $\sigma_{\mathrm{exp}} \lesssim 0.6~\mathrm{keV}$, corresponding to a precision better than $8\%$. Such precision is expected to be achievable at future collider experiments. At the High-Luminosity LHC, a sensitivity at the level of $\sim 14\%$ is anticipated~\cite{ATLASCMS2025}. Lepton colliders such as the ILC~\cite{ILC} are expected to reach a precision of $\sim 10\%$-$20\%$, enabling a robust test of new physics scenarios. Even higher precision could be achieved at FCC-ee~\cite{FCC}, where sensitivities at the level of $\sim 5\%$-$10\%$ are projected. These improvements would provide a powerful probe to distinguish RNS effects from the SM expectation. \\ While the current experimental sensitivity is insufficient to distinguish between those predictions, the parameter region that maximizes the $h \to Z\gamma$ decay is subject to stringent constraints from other Higgs observables. In particular, the $h \to \gamma\gamma$ channel, which is experimentally well measured and consistent with the SM, provides a critical consistency check. Any viable enhancement of the $h\to Z\gamma$ channel must therefore preserve agreement with the diphoton decay observables. 
\begin{table}[]
\centering
\begin{tabular}{|c|c|c|c|}
\hline
\textbf{Contribution} & \textbf{BP1} & \textbf{BP2} & \textbf{BP3} \\ \hline
\multicolumn{4}{|c|}{\texttt{ Main SM-like contributions}}\\ \hline
$g_{hWW}F_{1}(\tau_W)$   &  $-8.2822$  &  $-8.3452$   &  $-8.3975$    \\
$\frac{4}{3}g_{htt}F_{1/2}(\tau_t)$   &  $1.8336$    &  $1.8393$  &  $1.8502$    \\ 
$\frac{1}{3}g_{hbb}F_{1/2}(\tau_b)$ &  $-0.02575+0.03371\,i$   &  $-0.02475+0.03177\,i$    &  $-0.02719+0.03122\,i$    \\ \hline
\multicolumn{4}{|c|}{\texttt{ Main SUSY contributions}}\\ \hline
$g_{h\tilde{\chi}_1^+\tilde{\chi}_1^-} F_{1/2}(\tau_{\tilde \chi_1^+})$ & $-0.0231$ &  $-0.0225$   &  $-0.0216$    \\ 
$g_{h\tilde{\chi}_2^+\tilde{\chi}_2^-} F_{1/2}(\tau_{\tilde \chi_2^+})$ & $0.0112$     &  $0.0109$    & $9.95\times10^{-3}$ \\ 
$\frac{4}{3}g_{h\tilde t_1\tilde t_1}F_0(\tau_{\tilde t_1})$ & $-0.0171$ &   $-0.0180$  &  $-8.54\times10^{-3}$    \\ 
$\frac{4}{3}g_{h\tilde t_2\tilde t_2}F_0(\tau_{\tilde t_2})$ &  $0.0085$      &   $1.62\times10^{-3}$   &  $3.35\times10^{-4}$ \\  
$\frac{1}{3}g_{h\tilde b_1\tilde b_1}F_0(\tau_{\tilde b_1})$ & $-5.46\times10^{-5}$    &  $-3.36\times10^{-5}$    &  $-1.27\times10^{-5}$    \\
$g_{h\tilde \tau_1\tilde \tau_1}F_0(\tau_{\tilde \tau_1})$ & $-5.43\times10^{-4}$  &   $1.87\times10^{-4}$   &   $4.54\times10^{-5}$   \\ \hline
\multicolumn{4}{|c|}{\texttt{Total prediction}}\\ \hline
$\Gamma_{\rm \gamma\gamma}^{\rm RNS}$         &  $8.8059~\text{keV}$    &  $9.4589~\text{keV}$    &   $10.022~\text{keV}$   \\ \hline
\end{tabular}
\caption{\justifying \small{Individual one-loop contributions to the $h\rightarrow\gamma\gamma$ decay amplitude for three representative benchmark points. The quantities listed correspond to the main contributions entering the one-loop form factor, together with the resulting partial decay width.}}
\label{tab:hAA}
\end{table} 
The numerical results presented in Table~\ref{tab:hAA} allow us to identify the dominant contributions to the form factor $\mathcal{A}_{\rm \gamma\gamma}^{(2)}$ for the decay $h\rightarrow\gamma\gamma$ and to understand the origin of the differences among the three benchmark points defined in Table~\ref{tab:BP}. As expected, the amplitude is dominated by the SM contributions. The $W$-boson loop provides the largest contribution, with an amplitude varying from approximately $-8.28$ to $-8.40$, while the top-quark contribution remains close to $1.84$. These two contributions account for about $99\%$ of the total amplitude and interfere destructively, determining the overall magnitude of the decay width. The bottom-quark contribution is nearly two orders of magnitude smaller and mainly provides the small imaginary part of the amplitude. The supersymmetric contributions are considerably suppressed. The total SUSY correction to the real part of the amplitude remains below approximately $0.03$, corresponding to less than $0.5\%$ of the total amplitude. The light stop provides the dominant SUSY correction. Depending on the benchmark point, its contribution ranges from approximately $-0.018$ to $-0.009$, reflecting the increase of the stop mass from about $1.6$--$2.1~\rm{TeV}$. The heavy stop contribution is already one order of magnitude smaller because of the strong $1/m_{\tilde t}^2$ suppression entering the Higgs--squark coupling. The chargino contributions are of comparable magnitude, with the light chargino contributing about $-0.02$ and the heavy chargino approximately $+0.01$, leading to a partial cancellation. Finally, the sbottom and stau contributions are below $10^{-4}$ in the amplitude and can be regarded as completely negligible. It should be emphasized that the increase of the partial decay width is mainly driven by the Higgs-mass dependence appearing in the overall kinematic prefactor, $\Gamma(h\rightarrow\gamma\gamma)\propto M_h^3\,|\mathcal{A}_{\gamma\gamma}|^2$. Between BP1 and BP3 the Higgs mass increases from $123.78$ GeV to $127.72$ GeV, producing an enhancement of approximately $10\%$ through the cubic dependence on $M_h$. By comparison, the squared amplitude increases by about $3\%$.
\begin{figure*}[]
  \centering  
  \begin{subfigure}[c]{0.40\textwidth}
    \centering    \includegraphics[width=\linewidth]{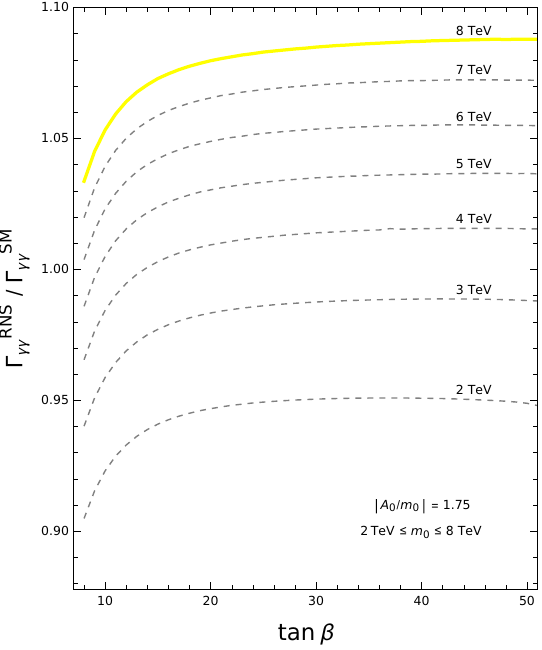}
    \caption{ }
    \label{fig:ratiophph}
  \end{subfigure} \hspace{1cm}
  \begin{subfigure}[c]{0.40\textwidth}
    \centering    \includegraphics[width=\linewidth]{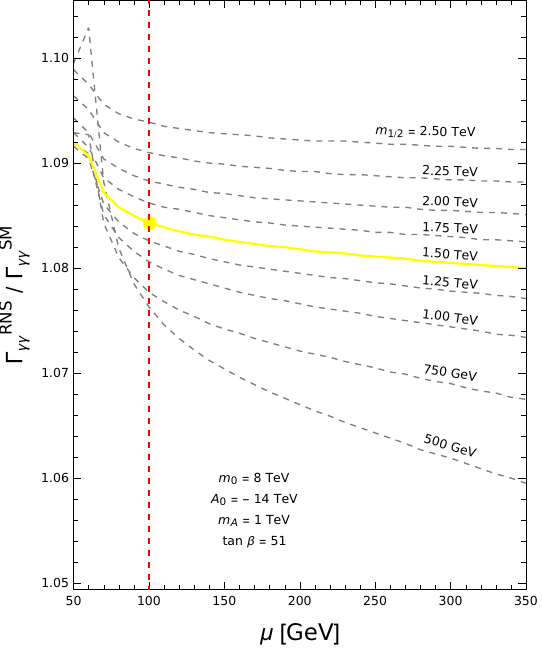}
        \caption{ }
        \label{fig:AAmum12}
  \end{subfigure}
   \caption{  \justifying \small{Diphoton decay properties of the Higgs boson in the RNS scenario. Panel (a) shows the ratio of the diphoton decay width in RNS to the SM prediction as a function of $\tan\beta$ for different values of the unified scalar mass $m_0$. Panel (b) shows the dependence of the normalized Higgs boson decay width into diphotons on the higgsino mass parameter $\mu$ for different values of the unified gaugino mass $m_{1/2}$ in the RNS scenario}}
    \label{fig:hgammagamma}
\end{figure*}\\
Figure~\ref{fig:ratiophph} shows the ratio of the $h\to \gamma\gamma$ decay width in RNS to its SM prediction as a function of $\tan\beta$, for different values of $m_0$ defined in (\ref{eq:RNSscan}), with $\mu=100~\text{GeV}$, $m_{1/2}=1.5~\text{TeV}$ and $m_A=1~\text{TeV}$. We observe that the diphoton decay width exhibits a mild dependence on $\tan\beta$, increasing at low values and approaching a plateau for $\tan\beta \gtrsim 20$. This behavior reflects the interplay between $\tan\beta$-dependent Higgs couplings and the decoupling of heavy superpartners entering the loop contributions. A more pronounced effect is observed in the dependence on $m_0$. For low values of $m_0 \sim 2~\mathrm{TeV}$, the diphoton rate is slightly suppressed relative to the SM prediction, while for larger values $m_0 \gtrsim 4~\mathrm{TeV}$, an enhancement is observed, reaching up to $\sim 8\%$ for $m_0 = 8~\mathrm{TeV}$. This transition originates from the interplay between heavy-stop radiative corrections and stop mixing, which governs the Higgs-boson mass prediction in the MSSM. The same interplay also affects the stop contribution to the form factor $\mathcal{A}_{\gamma \gamma}^{(2)}$, although its direct impact on the total diphoton amplitude remains subdominant. The yellow curve, corresponding to the parameter configuration that maximizes the $h \to Z\gamma$ decay width, lies in the region where the diphoton channel is also enhanced. This indicates a positive correlation between both decay modes, implying that an enhancement in $\Gamma_{Z\gamma}$ is accompanied by a moderate increase in the diphoton decay width. On the other hand, Figure~\ref{fig:AAmum12} shows the corresponding dependence of the diphoton partial decay width on the higgsino mass $\mu$ for several values of $m_{1/2}$. The vertical dashed red line indicates the minimum allowed value of $\mu$ ($\mu=100~{\rm GeV}$) according to current collider constraints, while the highlighted point corresponds to the RNS parameter set for which $\Gamma_{Z\gamma}$ reaches its maximum value of $7.5~\text{keV}$. The diphoton decay width exhibits only a weak sensitivity to the $\mu$ parameter. For the benchmark value $m_{1/2}=1.5~{\rm TeV}$, the variation of $\Gamma_{\gamma\gamma}^{\rm RNS}$ is below about $0.4\%$ throughout the interval $100~{\rm GeV}\leq\mu\leq350~{\rm GeV}$. Changing the unified gaugino mass over the range $500~{\rm GeV}\leq m_{1/2}\leq2.5~{\rm TeV}$ leads to small variations of approximately $3\%$, while preserving an enhancement of $\Gamma_{\gamma\gamma}^{\rm RNS}$ with respect to the SM prediction.
\begin{figure*}[]
  \centering  
  \begin{subfigure}[c]{0.40\textwidth}
    \centering    \includegraphics[width=\linewidth]{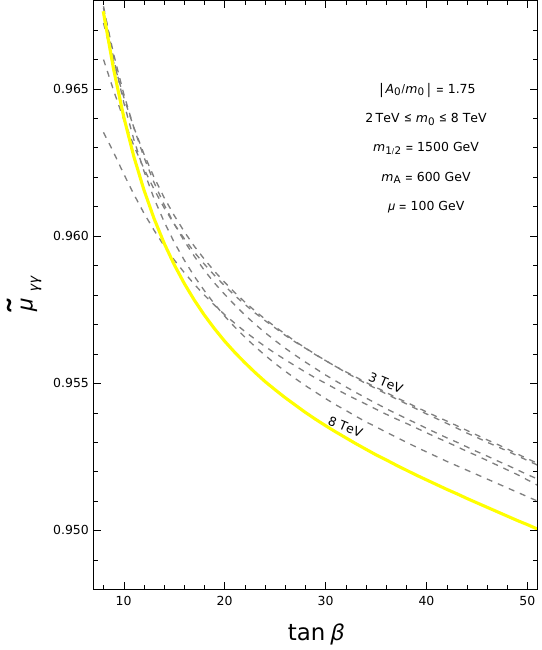}
    \caption{ }
    \label{fig:muphph}
  \end{subfigure} \hspace{1cm}
  \begin{subfigure}[c]{0.40\textwidth}
    \centering    \includegraphics[width=\linewidth]{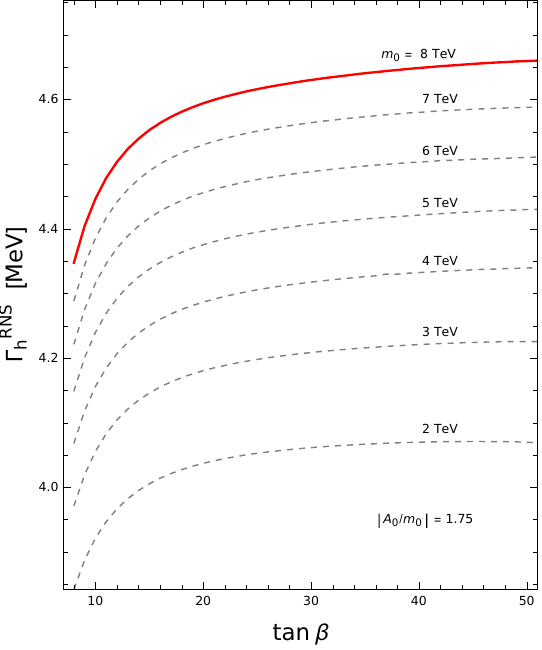}
        \caption{ }
        \label{fig:Gh}
  \end{subfigure}
   \caption{  \justifying \small{Panel (a) shows the RNS branching ratio normalized to the SM prediction, $\widetilde{\mu}_{\gamma\gamma}$, as a function of $\tan\beta$ for different values of the unified scalar mass $m_0$.} Panel (b) shows the total Higgs decay width in the RNS scenario as a function of $\tan\beta$ for different values of the unified scalar mass~$m_0$. Only decays into SM-like particles are included in $\Gamma_{\rm h}^{\text{RNS}}$.}
    \label{fig:mugammagamma}
\end{figure*}\\ \\ In Figure~\ref{fig:muphph}, we isolate the decay contribution to the Higgs signal strength by introducing the normalized branching ratio,
\begin{equation}
\widetilde{\mu}_{XY} = \frac{\mathrm{BR}(h\to XY)_{\rm RNS}}{\mathrm{BR}(h\to XY)_{\rm SM}} =
\frac{\mu_{XY}}{\sigma_{\rm RNS}/\sigma_{\rm SM}},
\label{eq:muXX}
\end{equation}
where $\mu_{XY}$ denotes the Higgs signal strength in the $XY$ final state. This definition explicitly removes the dependence on the Higgs production cross section, allowing us to study exclusively the effects of RNS radiative corrections on the Higgs decay branching ratios. For the diphoton channel we observe that $\widetilde{\mu}_{\gamma\gamma}$ is consistently below unity across the explored parameter space, even in regions where $\Gamma_{\gamma\gamma}$ is enhanced. This behavior is explained by the modification of the total Higgs decay width in RNS, which increases when $\tan \beta$ and $m_0$ increase, as can be shown in Figure~\ref{fig:Gh}. We have estimated $\Gamma_{\rm h}^{\rm RNS}$ using \texttt{FeynHiggs} 2.19.0~\cite{FeynHiggs}. The option \texttt{FHDECCZERO=1} was enabled in FeynHiggs to set numerically vanishing decay widths exactly to zero, thereby avoiding spurious contributions from numerical noise and mitigating the impact of numerical instabilities in the computation of branching ratios. In this setup, the Higgs boson decays only into SM-like particles; consequently, SUSY effects enter exclusively through loop corrections, resulting in a moderate variation of $\Gamma_{\rm h}^{\mathrm{RNS}}$ in the range $4.0-4.6~\mathrm{MeV}$. \\
As a result, the branching ratio into diphotons is reduced, leading to a suppression of the reduced signal strength. This effect becomes more pronounced at large $\tan\beta$, where $\widetilde{\mu}_{\gamma\gamma}$ decreases monotonically. The yellow curve, corresponding to the region that maximizes the $h \to Z\gamma$ decay width, lies within the same trend, indicating that an enhancement in $\Gamma_{Z\gamma}$ is accompanied by a mild suppression of $\widetilde{\mu}_{\gamma\gamma}$. The predicted values of $\widetilde{\mu}_{\gamma\gamma}$ remain close to the SM expectation, with $\widetilde{\mu}_{\gamma\gamma} \simeq 0.95-0.97$. These values lie well within current experimental measurements, which are at the level of $\widetilde{\mu}_{\gamma\gamma}^{\rm exp} \sim 1.0 \pm 0.1$~\cite{ATLASphph2023, ATLAS-CONF-2025-006}, indicating that the RNS parameter region that enhances the $h \to Z\gamma$ decay preserves the agreement with the precisely measured $h \to \gamma\gamma$ channel.  
\begin{table}[]
\centering
\begin{tabular}{|c|c|c|c|}
\hline
\textbf{Contribution} & \textbf{BP1} & \textbf{BP2} & \textbf{BP3} \\ \hline
\multicolumn{4}{|c|}{\texttt{ Main SM-like contributions}}\\ \hline
$g_{htt}F_{1/2}(\tau_t)$   &  $1.3752$    &  $1.3766$      &   $1.3799$   \\ 
$g_{hbb}F_{1/2}(\tau_b)$ & $-0.0774 + 0.1029\,i$   &   $-0.0732+0.0945\,i$   &  $-0.0712+0.0918\,i$    \\ \hline
\multicolumn{4}{|c|}{\texttt{ Main SUSY contributions}}\\ \hline
$g_{h\tilde t_1\tilde t_1}F_0(\tau_{\tilde t_1})$ & $-0.0126$ &  $- 0.0131$    &  $-0.0062$    \\ 
$g_{h\tilde t_2\tilde t_2}F_0(\tau_{\tilde t_2})$ & $2.737\times10^{-3}$      &  $1.391\times10^{-3}$    &  $3.22\times10^{-4}$   \\  
$g_{h\tilde b_1\tilde b_1}F_0(\tau_{\tilde b_1})$ & $-1.95\times10^{-4}$     &   $-1.18\times10^{-4}$   &  $-4.37\times10^{-5}$    \\
$g_{h\tilde b_2\tilde b_2}F_0(\tau_{\tilde b_2})$ & $-2.57\times10^{-6}$  &   $-8.84\times10^{-6}$   &  $-4.30\times10^{-6}$    \\ \hline
\multicolumn{4}{|c|}{\texttt{Total prediction}}\\ \hline
$\Gamma_{\rm gg}^{\rm RNS}$         &  $0.2803~\text{MeV}$    &  $0.2924~\text{MeV}$    &   $0.3069~\text{MeV}$   \\ \hline
\end{tabular}
\caption{\justifying \small{Individual contributions to the $h\rightarrow gg$ decay amplitude for the three representative benchmark points defined in Table~\ref{tab:BP}. The quantities listed correspond to the main contributions entering the one-loop form factor, together with the resulting partial decay width.}}
\label{tab:Agg}
\end{table} \\ \\
In addition to the diphoton channel, it is instructive to analyze the Higgs decay into two gluons in the same region of parameter space. Table~\ref{tab:Agg} shows the main contributions to $\Gamma_{\rm gg}^{\rm RNS}$ for the benchmark points listed in Table~\ref{tab:BP}. The numerical evaluation of the individual loop corrections reveals a clear hierarchy among the contributions entering the $h\rightarrow gg$ amplitude. In all three benchmark points, the top-quark loop provides the dominant contribution, with a nearly constant amplitude of approximately $1.38$. The bottom-quark contribution is smaller, with a magnitude of about $7\times10^{-2}$ and a non-negligible imaginary part originating from the condition $\tau_b<1$. Although this contribution is roughly twenty times smaller than the top amplitude, it interferes destructively with the real part of the top loop and therefore cannot be neglected. Among the SUSY particles, the light stop $\tilde t_1$ is the only state that gives a relevant correction to the amplitude. In the MSSM, the Higgs-stop coupling $g_{h\tilde t_1 \tilde t_1}$ contains a mixing term proportional to $\sin(2\theta_{\tilde t})\,\tilde{A}_t/m_{\tilde t_1}^2$, which can be substantially enhanced for sufficiently large values of the effective mixing parameter $\tilde{A}_t$ together with nearly maximal stop mixing ($\sin2\theta_{\tilde t}\simeq1$). In this case, $g_{h\tilde t_1 \tilde t_1}$ may become comparable to or even larger than the Higgs-top coupling $g_{htt}$, leading to sizeable SUSY corrections to the $h\rightarrow gg$ decay amplitude~\cite{Djouadi1998}. However, the RNS benchmarks considered in this work exhibit a markedly different behavior. Although the effective mixing parameter $|\tilde{A}_t|$ increases significantly from BP1 to BP3, the stop mixing angle decreases simultaneously, while the stop masses remain at the TeV scale. As a consequence, the enhancement expected from larger values of $|\tilde{A}_t|$ is largely compensated by the suppression associated with the smaller factor $\sin2\theta_{\tilde t}$ and the heavier stop masses. The dominant mixing contribution therefore remains moderate, and the light-stop correction stays at the percent level throughout the three benchmark points. This behavior reflects the characteristic spectrum generated by the RNS boundary conditions, which favor a region of the MSSM parameter space where large stop mixing and light stops do not occur simultaneously, preventing large enhancements. Finally, the sbottom sector is found to be subdominant in all benchmark points. Both $\tilde b_1$ and $\tilde b_2$ contribute at the level of $10^{-4}$--$10^{-5}$ to the total amplitude. Overall, the total SUSY contribution modifies the SM-like amplitude by about $2\%$, depending on the benchmark point in the RNS scenario. It is worth noting that for BP2, which predicts $M_h\simeq125~\rm{GeV}$, the RNS partial width is $\Gamma_{\rm gg}^{\rm RNS}=0.2924~\text{MeV}$. Although the central value of the RNS prediction corresponds to approximately $85\%$ of the SM prediction reported by the PDG for a Higgs boson of the same mass, $\Gamma^{\rm SM}_{\rm gg}\simeq0.34\pm 0.025~\text{MeV}$, this fraction varies between about $80\%$ and $93\%$ once the uncertainty is taken into account. The relatively wide interval is derived from the $\sim7.2\%$ uncertainty in the predicted branching ratio $BR(h\to gg)$, and the uncertainty in the total Higgs width, which amounts to $\sim 1.37\%$.
\begin{figure*}[]
  \centering  
  \begin{subfigure}[c]{0.40\textwidth}
    \centering    \includegraphics[width=\linewidth]{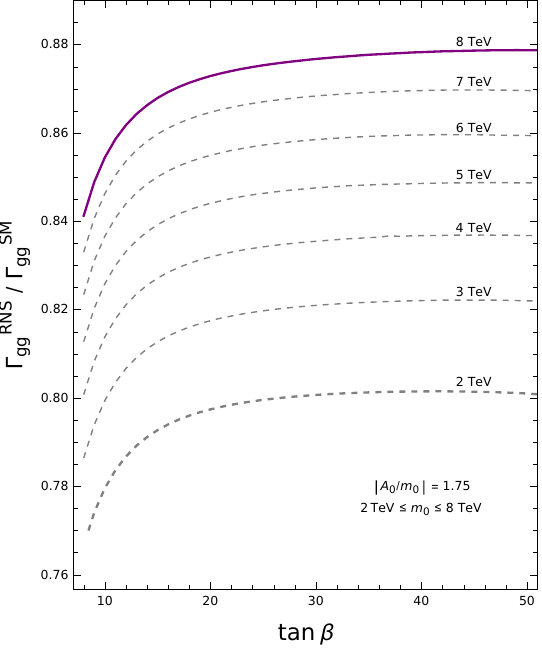}
    \caption{ }
    \label{fig:hgg}
  \end{subfigure} \hspace{1cm}
  \begin{subfigure}[c]{0.40\textwidth}
    \centering    \includegraphics[width=\linewidth]{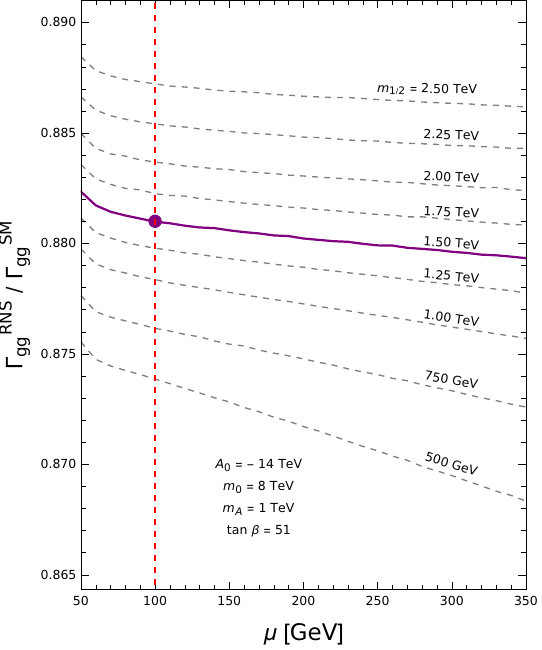}
        \caption{}
        \label{fig:ggmum12}
  \end{subfigure}
   \caption{  \justifying \small{(a) Ratio of the Higgs decay width into gluons in the RNS scenario to the SM prediction as a function of $\tan\beta$ for different values of $m_0$. (b) Dependence of the normalized Higgs boson decay width into gluons on the higgsino mass parameter $\mu$ for different values of the unified gaugino mass $m_{1/2}$ in the RNS scenario.}}
    \label{fig:Allmodes}
\end{figure*} \\
Additionally, Figure~\ref{fig:hgg} shows the ratio of the $h \to gg$ decay width in RNS to the SM prediction as a function of $\tan\beta$ for different values of $m_0$. The dependence on $m_0$ reflects the enhanced sensitivity of $\Gamma_{\rm gg}$ to the squark loop corrections, particularly from the stop sector, which is the dominant SUSY contribution. Using the central PDG value as a reference, we find that the partial width is systematically suppressed, with $\Gamma_{gg}^{\mathrm{RNS}} / \Gamma_{gg}^{\mathrm{SM}} \simeq 0.78-0.88$, implying a reduction of approximately $10\%-20\%$. The purple solid line, corresponding to the parameter configurations that maximize the $h \to Z\gamma$ decay width, exhibit a non-negligible deviation from the SM prediction. In particular, at the point that maximizes $\Gamma_{Z\gamma}$, the deviation in $\Gamma_{gg}$ is at the level of $\sim 10\%$, moderately larger than the $\sim 5\%$ deviations observed in the diphoton channel. While such deviations seem to introduce some tension with current experimental constraints, it is important to note that current uncertainties in the branching ratio BR($h\to gg$), as well as in the $hgg$ effective coupling, still allow for $\mathcal{O}(10\%)$ deviations~\cite{PDG2026, deFlorian2016}. Nevertheless, these results highlight a trade-off in the parameter space, regions that maximize the $h \to Z\gamma$ decay may induce sizable deviations in other Higgs observables, providing complementary probes of the RNS scenario. \\
Figure~\ref{fig:ggmum12} shows the ratio $\Gamma_{gg}^{\rm RNS}/\Gamma_{gg}^{\rm SM}$ as a function of the higgsino mass $\mu$ for different values of the unified gaugino mass $m_{1/2}$. The vertical dashed red line indicates again the reference value $\mu=100~{\rm GeV}$, while the highlighted point corresponds to the benchmark scenario where $\Gamma_{Z\gamma}$ is maximized. The figure shows that the gluonic partial width exhibits only a mild dependence on $\mu$. For the value $m_{1/2}=1.5~{\rm TeV}$, the variation of $\Gamma_{gg}^{\rm RNS}$ remains below approximately $0.2\%$ over the phenomenologically relevant interval $100\leq\mu\leq350~{\rm GeV}$. Varying $m_{1/2}$ between $500~{\rm GeV}$ and $2.5~{\rm TeV}$ produces a somewhat larger but still moderate shift of about $2\%$. These results indicate that the suppression of the gluonic decay width predicted in the RNS scenario is stable against variations of both $\mu$ and $m_{1/2}$ within the considered parameter space.
\begin{figure}[]
    \centering
    \includegraphics[width=0.47\textwidth]{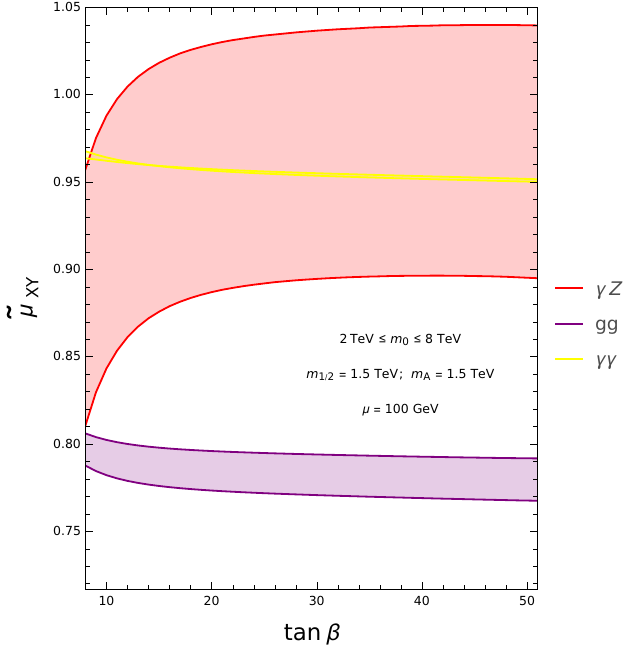}
    \caption{ 
        \justifying \small{RNS branching ratios normalized to the SM predictions of the loop-induced Higgs decay channels $h \to Z\gamma$ (red band), $h \to \gamma\gamma$ (yellow band), and $h \to gg$ (purple band) in the RNS scenario as functions of $\tan\beta$ for different values of $m_0$.}
       }
    \label{fig:Gauge}
\end{figure} \\ \\  
In Figure~\ref{fig:Gauge}, we present a global view of $\widetilde{\mu}_{XY}$ for all loop-induced Higgs decay channels in the RNS scenario. The bands reflect the variation of the parameter $m_0$ in the range $2~\mathrm{TeV} \le m_0 \le 8~\mathrm{TeV}$, while $\tan\beta$ is varied between $8$ and $51$ for $m_A = 1.5~\mathrm{TeV}$ and $\mu = 100~\mathrm{GeV}$. It is important to note that the gluon channel (purple curve) can exhibit sizable suppressions, with $\widetilde{\mu}_{gg}\simeq 0.8$, corresponding to deviations of order $20\%$ relative to the SM prediction. However, the effective Higgs--gluon--gluon coupling governs not only the decay $h\to gg$, but also the dominant Higgs production mechanism through gluon fusion. Consequently, a direct comparison with the experimentally measured Higgs signal strengths requires the corresponding modifications to the Higgs production cross sections to be taken into account, which lies beyond the scope of this work. Nevertheless, the predicted branching-ratio modifications provide a useful characterization of the dependence of the gluon channel on $m_0$ and $\tan\beta$, while allowing a direct comparison with the behavior of the other loop-induced Higgs decay modes. \\ 
A common feature of the $\gamma\gamma$, $Z\gamma$, and $gg$ channels is their weak dependence on $\tan\beta \gtrsim 20$. In contrast, this channels exhibit a more pronounced dependence at low $\tan\beta \lesssim 20$, indicating that radiative corrections included in the form factors $\mathcal{A}_{V_1 V_2}^{(2)}$ have a strong dependence on $\tan \beta$ in this regime. The width of each band reflects the dependence on the scalar mass parameter $m_0$. In particular, the $Z\gamma$ channel displays the largest spread, with $\widetilde{\mu}_{Z\gamma}$ varying approximately from $0.85$ to $1.05$, corresponding to enhancements of up to about $20\%$ relative to the SM expectation. By contrast, the gluon channel remains systematically suppressed, with $\widetilde{\mu}_{gg}$ changing only slightly from about $0.78$ to $0.82$, corresponding to a variation of roughly $5\%$ within the scan, while staying about $18-22\%$ below the SM prediction. The diphoton channel shows an even weaker dependence on $m_0$, with $\widetilde{\mu}_{\gamma\gamma}$ lying in the narrow interval $\sim 0.95-0.97$, corresponding to only a $2\%$ variation across the scan and deviations of at most $3-5\%$ from the SM value. Overall, Figure~\ref{fig:Gauge} highlights the complementary behavior of the Higgs decays into gauge bosons within the RNS framework. While the $h \to Z\gamma$ decay can be significantly enhanced, the diphoton channel remains consistent with current experimental constraints. The gluon channel, on the other hand, provides a particularly sensitive probe of SUSY effects.

\vspace*{-0.2cm}
\section{\label{sec:conclusions} Conclusions}

In this work, we have studied the loop-induced Higgs decay channels $h \to Z\gamma$, $h \to \gamma\gamma$, and $h \to gg$ within the framework of Radiative Natural Supersymmetry. As a first step, we reproduced the one-loop MSSM predictions for the corresponding partial decay widths, obtaining a consistent analytical and numerical description of these processes in the RNS scenario. This allowed us to perform a dedicated scan of the parameter region where the rare decay $h \to Z\gamma$ is maximized. Our analysis shows that the $h \to Z\gamma$ channel receive a significant enhancement in RNS. In the preferred region of parameter space, the decay width reaches a maximum value of approximately $7.5~\mathrm{keV}$, compared with the SM prediction $\Gamma_{Z\gamma}^{\rm SM}=6.19 \pm 0.15~\mathrm{keV}$, corresponding to an increase of about $20\%$. This prediction remains compatible with the latest combined ATLAS result, although the present experimental uncertainty is still too large to discriminate between the SM and the RNS scenario. We also examined the impact of the same parameter region on the other loop-induced Higgs decays. The diphoton channel remains remarkably stable, with a branching ratio normalized to the SM prediction in the range $\widetilde{\mu}_{\gamma\gamma}\simeq 0.95-0.97$, corresponding to deviations below $5\%$ from the SM expectation. This result is particularly relevant, since the $h \to \gamma\gamma$ mode is one of the most precisely measured Higgs channels at the LHC, and therefore provides a stringent consistency test of the scenario. The $h \to gg$ channel exhibits a stronger sensitivity to the SUSY corrections present in the selected RNS parameter region. We find a suppression of the partial decay width by about $10\%$ relative to the SM prediction, leading to a normalized branching ratio of approximately $\widetilde{\mu}_{gg}\simeq 0.8$. This behavior indicates that the effective Higgs-gluon-gluon coupling can provide an additional indirect probe of SUSY effects, a fully consistent phenomenological treatment requires the inclusion of modified gluon-fusion production rates. The dependence of the observables on the parameters $m_0$ and $\tan\beta$ reveals characteristic correlations among the three loop-induced channels. In particular, the $Z\gamma$ mode shows the strongest sensitivity to variations in $m_0$,  while the diphoton channel exhibits only negligible changes throughout the scanned $m_0$ range. At large $\tan\beta$, most observables tend to stabilize, whereas larger variations may appear in the low-$\tan\beta$ region. In summary, our results show that RNS can enhance the rare decay $h \to Z\gamma$ while preserving consistency with current constraints from the $h \to \gamma\gamma$ channel and inducing testable deviations in the $h \to gg$ decay width. Future precision measurements of the $h \to Z\gamma$ decay width at the HL-LHC and future Higgs factories will therefore play an important role in probing this SUSY scenario.

\begin{acknowledgments}
This work was partially supported by the research grant \textit{SIGP 400-156.012-014 (GA313-BP-2024) Observables de alta precisión en la física del bosón de Higgs}, from the call \textit{Convocatoria Interna de Banco de Proyectos - Año 2024 - Universidad de Pamplona}.
\end{acknowledgments}

\appendix

\section{\label{sec:ap_A} Higgs Couplings in the MSSM}

In this appendix we summarize the structure of the relevant Higgs couplings entering the decay amplitudes. These couplings depend on the Higgs mixing angles $\alpha$ and $\beta$, the weak isospin $T_3$, the electric charge $Q_i$, as well as on SUSY parameters such as the Higgsino mass $\mu$, the trilinear soft breaking parameter $A_f$, and the sfermion mixing matrices. The reduced couplings of the light CP-even Higgs boson to fermions are given by
\begin{align}
g_{huu} &=  \frac{\cos\alpha}{\sin\beta}, \\
g_{hdd} &= - \frac{\sin\alpha}{\cos\beta},
\end{align}
The reduced Higgs--sfermion couplings in the sfermion mass basis are given by
\begin{align}
g_{h\tilde{f}_1\tilde{f}_1} &=
-\frac{M_Z^2}{m_{\tilde{f}_1}^2}\kappa_V \cos(2\beta)
\left(
I_3^f \cos^2\theta_{\tilde f}
-
Q_f s_W^2 \cos2\theta_{\tilde f}
\right)
-\kappa_f\,\frac{m_f^2}{m_{\tilde{f}_1}^2}
+\frac12
\sin2\theta_{\tilde f}
\frac{m_f \tilde{A}_f}{m_{\tilde{f}_1}^2},
\\[2mm]
g_{h\tilde{f}_2\tilde{f}_2} &=
-\frac{M_Z^2}{m_{\tilde{f}_2}^2}\kappa_V \cos(2\beta)
\left(
I_3^f \sin^2\theta_{\tilde f}
+
Q_f s_W^2 \cos2\theta_{\tilde f}
\right)
-\kappa_f\,\frac{m_f^2}{m_{\tilde{f}_2}^2}
-\frac12
\sin2\theta_{\tilde f}
\frac{m_f\tilde{A}_f}{m_{\tilde{f}_2}^2},
\end{align}
where $\tilde{A}_f = A_f\kappa_f-\mu\kappa_f'$ and
\begin{align}
\kappa_u &= \frac{\cos\alpha}{\sin\beta},
&
\kappa_d &= -\frac{\sin\alpha}{\cos\beta},
\\
\kappa_u' &= \cot\beta,
&
\kappa_d' &= \tan\beta,
\\
\kappa_V &= \sin(\beta-\alpha).
\end{align}
The Higgs couplings to charginos are given by
\begin{equation}
g_{h\tilde{\chi}_j^+\tilde{\chi}_j^-} = - \frac{\sqrt{2}M_W}{m_{\tilde{\chi}_j}} \left( V_{j1} U_{j2} \sin\alpha - V_{j2} U_{j1} \cos\alpha \right),
\end{equation}
where $U$ and $V$ are the chargino mixing matrices. The Higgs couplings to electroweak $W$ bosons take the form
\begin{equation}
g_{hWW} = \sin(\beta - \alpha).
\end{equation}
The reduced couplings of the $Z$ boson entering the $h\to Z\gamma$ decay amplitude are given by
\begin{align}
g_{ZWW} &= c_W ,
\\[2mm]
g_{Zff} &= \frac{T_3^f-2Q_f s_W^2}{c_W},
\\[2mm]
g_{Z\chi_j^+\chi_j^-}
& =
\frac{1}{2}
\left(
V_{i1}V_{j1} + \frac{1}{2}V_{i2}V_{j2} -\delta_{ij}s_W^2 +
U_{i1}U_{j1} + \frac{1}{2}U_{i2}U_{j2} -\delta_{ij}s_W^2
\right),
\\[2mm]
g_{Z\tilde f_i\tilde f_j}
&=
\frac{1}{c_W}
\left(
T_3^f R^{\tilde f}_{i1}R^{\tilde f}_{j1}
-
Q_f s_W^2\delta_{ij}
\right),
\end{align}
where $R^{\tilde f}$ is the sfermion mixing matrix. The full expressions for these couplings, including all mixing effects and sign conventions, are implemented in detail in \texttt{FeynArts} and can be obtained directly from the MSSM model files.

\nocite{*}

\bibliography{apssamp}

@PREAMBLE{
 "\providecommand{\noopsort}[1]{}" 
 # "\providecommand{\singleletter}[1]{#1}%" 
}

@article{2023HiggsSM,
  author       = "Reyes R., Edilson A. and Fazio, Angelo R.",
  title        = {},
  journal      = {Phys. Rev. D},
  volume       = {108},
  number       = {5},
  pages        = {053007},
  year         = {2023},
  doi          = {10.1103/PhysRevD.108.053007},
  url          = {https://doi.org/10.1103/PhysRevD.108.053007}
}

@article{ATLASCMS,
  author = {Aad, G. and others},
  collaboration = {ATLAS Collaboration and CMS Collaboration},
  title   = {},
  journal = {J. High Energy Phys.},
  volume  = {2016},
  number  = {45},
  pages   = {08},
  year    = {2016},
  doi     = {10.1007/JHEP08(2016)045},
  url     = {https://link.springer.com/article/10.1007/JHEP08(2016)045}
}

@article{ATLAS2023,
  author  = {Aad, Georges and others},
  title   = {},
  collaboration = {ATLAS},
  journal = {Phys. Lett. B},
  volume  = {847},
  pages  = {138315},
  year    = {2023},
  doi     = {10.1016/j.physletb.2023.138315}
}

@article{CMS2025,
  author  = {Hayrapetyan, Armen and others},
  title   = {},
  collaboration = {CMS},
  journal = {Phys. Rev. D},
  volume  = {111},
  number  = {9},
  pages   = {092014},
  year    = {2025},
  doi     = {10.1103/PhysRevD.111.092014}
}

@article{Martin2019,
  author  = "Martin, Stephen P. and Robertson, David G.",
  title   = {},
  journal = {Phys. Rev. D},
  volume  = {100},
  number  = {7},
  pages   = {073004},
  year    = {2019},
  doi     = {10.1103/PhysRevD.100.073004}
}

@article{Martin2021,
  author  = "Martin, Stephen P.",
  title   = {},
  journal = {Phys. Rev. D},
  volume  = {105},
  number  = {5},
  pages   = {056014},
  year    = {2022},
  doi     = {10.1103/PhysRevD.105.056014}
}

@article{Martin2022,
  author  = "Martin, Stephen P.",
  title   = {},
  journal = {Phys. Rev. D},
  volume  = {106},
  number  = {1},
  pages   = {013007},
  year    = {2022},
  doi     = {10.1103/PhysRevD.106.013007}
}

@article{Martin2023,
  author = "Martin, Stephen P.",
  title   = {},
  journal = {Phys. Rev. D},
  volume = {107},
  issue = {5},
  pages = {053005},
  numpages = {27},
  year = {2023},
  month = {Mar},
  publisher = {American Physical Society},
  doi = {10.1103/PhysRevD.107.053005},
  url = {https://link.aps.org/doi/10.1103/PhysRevD.107.053005}
}

@article{2022Particles,
  author       = "Reyes R., Edilson A. and Fazio, Angelo R.",
  title        = {},
  journal      = {Particles},
  volume       = {5},
  number       = {1},
  pages        = {53-73},
  year         = {2022},
  doi          = {10.3390/particles5010006},
  url          = {https://doi.org/10.3390/particles5010006}
}

@article{2021EPJC,
  author       = "Pietro Slavich and others",
  title        = {},
  journal      = {Eur. Phys. J. C},
  volume       = {81},
  number       = {5},
  pages        = {450},
  year         = {2021},
  doi          = {10.1140/epjc/s10052-021-09198-2},
  url          = {https://doi.org/10.1140/epjc/s10052-021-09198-2}
}

@article{Djouadi2007,
author = {Abdelhak Djouadi},
title = {},
journal = {Phys. Rept.},
volume = {457},
number = {1},
pages = {1-216},
year = {2008},
issn = {0370-1573},
doi = {https://doi.org/10.1016/j.physrep.2007.10.004},
url = {https://www.sciencedirect.com/science/article/pii/S0370157307004334}
}

@article{Spira2017,
  author  = {Spira, Michael},
  title   = {},
  journal = {Prog. Part. Nucl. Phys.},
  volume  = {95},
  pages   = {98--159},
  year    = {2017},
  doi     = {10.1016/j.ppnp.2017.04.001},
  url     = {https://doi.org/10.1016/j.ppnp.2017.04.001}
}

@article{Djouadi2008,
  author  = {Djouadi, Abdelhak},
  title   = {},
  journal = {Phys. Rept.},
  volume  = {459},
  pages   = {1--241},
  year    = {2008},
  doi     = {10.1016/j.physrep.2007.10.005},
  url     = {https://doi.org/10.1016/j.physrep.2007.10.005},
  }

@article{CarenaHaber2003,
author = {M. Carena and H.E. Haber},
title = {},
journal = {Prog. Part. Nucl. Phys.},
volume = {50},
number = {1},
pages = {63-152},
year = {2003},
issn = {0146-6410},
doi = {https://doi.org/10.1016/S0146-6410(02)00177-1},
url = {https://www.sciencedirect.com/science/article/pii/S0146641002001771}
}

@article{HiggsExp1,
  author = {Aad, G. and others},
  collaboration = {ATLAS Collaboration and CMS Collaboration},
  title   = {},
  journal = {J. High Energy Phys.},
  volume  = {08},
  pages   = {045},
  year    = {2016},
  doi     = {10.1007/JHEP08(2016)045},
  url     = {https://link.springer.com/article/10.1007/JHEP08(2016)045}
}

@article{HiggsExp2,
  author  = {CMS Collaboration},
  title   = {},
  journal = {Eur. Phys. J. C},
  volume  = {79},
  number  = {5},
  pages   = {421},
  year    = {2019},
  doi     = {10.1140/epjc/s10052-019-6909-y}
  }

@article{hgZcomb,
  title = {},
  author = {Aad, G. and others},
  collaboration = {ATLAS Collaboration and CMS Collaboration},
  journal = {Phys. Rev. Lett.},
  volume = {132},
  issue = {2},
  pages = {021803},
  numpages = {32},
  year = {2024},
  month = {Jan},
  publisher = {American Physical Society},
  doi = {10.1103/PhysRevLett.132.021803},
  url = {https://link.aps.org/doi/10.1103/PhysRevLett.132.021803}
}

@article{hgZAtlas,
    author = "Aad, Georges and others",
    collaboration = "ATLAS",
    journal = {},
    title = "{Search for the Higgs boson decay to a $Z$ boson and a photon in $pp$ collisions at $\sqrt{s}=13$ TeV and $13.6$ TeV with the ATLAS detector}",
    eprint = "2507.12598",
    archivePrefix = "arXiv",
    primaryClass = "hep-ex",
    reportNumber = "CERN-EP-2025-155",
    month = "7",
    year = "2025"
}

@inproceedings{ILC,
    author = "Asner, D. M. and others",
    title = "{ILC Higgs White Paper}",
    booktitle = "{Snowmass 2013}: {Snowmass on the Mississippi}",
    eprint = "1310.0763",
    archivePrefix = "arXiv",
    primaryClass = "hep-ph",
    month = "10",
    year = "2013"
}

@article{FCC,
  author  = {Abada, A. and Abbrescia, M. and Abdus Salam, S. S. and others},
  title   = {FCC Physics Opportunities},
  journal = {Eur. Phys. J. Special Topics},
  volume  = {228},
  pages   = {261--623},
  year    = {2019},
  doi     = {10.1140/epjst/e2019-900045-4},
  url     = {https://doi.org/10.1140/epjst/e2019-900045-4}
}

@article{Baer2012,
  title = {},
  author = {Baer, Howard and Barger, Vernon and Huang, Peisi and Mustafayev, Azar and Tata, Xerxes},
  journal = {Phys. Rev. Lett.},
  volume = {109},
  issue = {16},
  pages = {161802},
  numpages = {5},
  year = {2012},
  month = {Oct},
  publisher = {American Physical Society},
  doi = {10.1103/PhysRevLett.109.161802},
  url = {https://link.aps.org/doi/10.1103/PhysRevLett.109.161802}
}

@article{Baer2013,
  title = {},
  author = {Baer, Howard and Barger, Vernon and Huang, Peisi and Mickelson, Dan and Mustafayev, Azar and Tata, Xerxes},
  journal = {Phys. Rev. D},
  volume = {87},
  issue = {11},
  pages = {115028},
  numpages = {21},
  year = {2013},
  month = {Jun},
  publisher = {American Physical Society},
  doi = {10.1103/PhysRevD.87.115028},
  url = {https://link.aps.org/doi/10.1103/PhysRevD.87.115028}
}

@article{Baer2022,
  author  = {Baer, Howard A. and Barger, Vernon and Martinez, Dakotah and Salam, Shadman},
  title   = {},
  journal = {J. High Energy Phys.},
  volume  = {2022},
  number  = {3},
  pages   = {186},
  year    = {2022},
  doi     = {10.1007/JHEP03(2022)186},
  url     = {https://doi.org/10.1007/JHEP03(2022)186}
}

@article{Edilson2025,
  author = {Reyes R., E. A. and Lopez A., C. A. and Torrijo G., O. R. and Melo P., D. G.},
   title = {},
  journal = {Phys. Rev. D},
  volume = {112},
  issue = {9},
  pages = {095012},
  numpages = {12},
  year = {2025},
  month = {Nov},
  publisher = {American Physical Society},
  doi = {10.1103/q9j9-2zkq},
  url = {https://link.aps.org/doi/10.1103/q9j9-2zkq}
}

@article{ATLAS2025,
author = {Aad, G. and others},
title = {},
journal = {Physics Letters B},
pages = {140313},
year = {2026},
issn = {0370-2693},
doi = {https://doi.org/10.1016/j.physletb.2026.140313},
url = {https://www.sciencedirect.com/science/article/pii/S037026932600167X}
}

@techreport{CMS2026,
      author = {},
      collaboration = "CMS",
      title         = "{Search for the rare Higgs boson decay H to Z gamma in proton-proton collisions at sqrt(s)=13 and 13.6 TeV}",
      institution   = "CERN",
      reportNumber  = "CMS-PAS-HIG-25-010",
      address       = "Geneva",
      year          = "2026",
      url           = "https://cds.cern.ch/record/2958406",
}

@misc{data,
  author       = {Reyes R., E. A.},
  title        = {Ancillary files for ``Loop-Induced Higgs Boson Decays into Gauge Bosons in Radiative Natural Supersymmetry''},
  year         = {2026},
  howpublished = {\url{https://github.com/fisicateoricaUDP/HiggsDecays.git}}
}

@article{Isasugra,
  title = {},
  author = {Baer, H. and Ferrandis, J. and Kraml, S. and Porod, W.},
  journal = {Phys. Rev. D},
  volume = {73},
  issue = {1},
  pages = {015010},
  numpages = {9},
  year = {2006},
  month = {Jan},
  publisher = {American Physical Society},
  doi = {10.1103/PhysRevD.73.015010},
  url = {https://link.aps.org/doi/10.1103/PhysRevD.73.015010}
}

@article{Isajet,
    author = "Paige, Frank E. and Protopopescu, Serban D. and Baer, Howard and Tata, Xerxes",
    title = "{ISAJET 7.69: A Monte Carlo event generator for pp, anti-p p, and e+e- reactions}",
    journal = {},
    eprint = "hep-ph/0312045",
    archivePrefix = "arXiv",
    month = "12",
    year = "2003"
}

@article{DRED,
title = {},
journal = {Physics Letters B},
volume = {84},
number = {2},
pages = {193-196},
year = {1979},
issn = {0370-2693},
doi = {https://doi.org/10.1016/0370-2693(79)90282-X},
url = {https://www.sciencedirect.com/science/article/pii/037026937990282X},
author = {Warren Siegel}
}

@article{CAPPER,
title = {},
journal = {Nuclear Physics B},
volume = {167},
number = {3},
pages = {479-499},
year = {1980},
issn = {0550-3213},
doi = {https://doi.org/10.1016/0550-3213(80)90244-8},
url = {https://www.sciencedirect.com/science/article/pii/0550321380902448},
author = {D.M. Capper and D.R.T. Jones and P. {Van Nieuwenhuizen}}
}

@article{Dominik2005,
doi = {10.1088/1126-6708/2005/03/076},
url = {https://doi.org/10.1088/1126-6708/2005/03/076},
year = {2005},
month = {apr},
publisher = {},
volume = {2005},
number = {03},
pages = {076},
author = {Dominik Stöckinger},
title = {},
journal = {Journal of High Energy Physics}
}

@article{PDG2026,
    author = "Takahashi, F. and others",
    collaboration = "Particle Data Group",
    title = "{Review of Particle Physics}",
    doi = "10.1142/S0217751X26300115",
    url = {https://pdg.lbl.gov/},
    journal = "Int. J. Mod. Phys. A",
    volume = "41",
    pages = "2630011",
    year = "2026"
}

@article{Dobado,
  author       = {Dobado, Antonio and Herrero, Maria J. and Pe\~naranda, Silvia},
  title        = {},
  journal      = {Eur. Phys. J. C},
  volume       = {17},
  pages        = {487--499},
  year         = {2000},
  doi          = {https://doi.org/10.1007/s100520000486},
  url          = {https://link.springer.com/article/10.1007/s100520000486#citeas}
}

@article{Ellis1976,
  author       = "Ellis, John R. and Gaillard, Mary K. and Nanopoulos, Dimitri V.",
  title        = {},
  journal      = {Nuclear Physics B},
  volume       = {106},
  pages        = {292},
  year         = {1976},
  doi          = {10.1016/0550-3213(76)90382-5}
}

@article{Shifman2012,
  title = {},
  author = {Shifman, M. and Vainshtein, A. and Voloshin, M. B. and Zakharov, V.},
  journal = {Phys. Rev. D},
  volume = {85},
  issue = {1},
  pages = {013015},
  numpages = {4},
  year = {2012},
  month = {Jan},
  publisher = {American Physical Society},
  doi = {10.1103/PhysRevD.85.013015},
  url = {https://link.aps.org/doi/10.1103/PhysRevD.85.013015}
}

@article{Cahn1979,
  author       = "Cahn, R. N. and Chanowitz, M. S. and Fleishon, N.",
  title        = {},
  journal      = {Phys. Lett. B},
  volume       = {82},
  pages        = {113--116},
  year         = {1979},
  doi          = {10.1016/0370-2693(79)90438-6}
}

@article{Bergstrom1985,
  author       = "Bergstr{\"o}m, L. and Hulth, G.",
  title        = {},
  journal      = {Nuclear Physics B},
  volume       = {259},
  pages        = {137--155},
  year         = {1985},
  doi          = {10.1016/0550-3213(85)90302-5}
}

@article{Spira1992,
title = {},
journal = {Physics Letters B},
volume = {276},
number = {3},
pages = {350-353},
year = {1992},
issn = {0370-2693},
doi = {https://doi.org/10.1016/0370-2693(92)90331-W},
url = {https://www.sciencedirect.com/science/article/pii/037026939290331W},
author = {M. Spira and A. Djouadi and P.M. Zerwas}
}

@article{Gehrmann2015,
  author    = {T. Gehrmann and S. Guns and D. Kara},
  title     = {},
  journal   = {Journal of High Energy Physics},
  year      = {2015},
  volume    = {2015},
  number    = {9},
  pages     = {038},
  doi       = {10.1007/JHEP09(2015)038},
  url = {https://link.springer.com/article/10.1007/JHEP09(2015)038#citeas}
}

@article{Chen2024,
  title = {},
  author = {Chen, Zi-Qiang and Chen, Long-Bin and Qiao, Cong-Feng and Zhu, Ruilin},
  journal = {Phys. Rev. D},
  volume = {110},
  issue = {5},
  pages = {L051301},
  numpages = {5},
  year = {2024},
  month = {Sep},
  publisher = {American Physical Society},
  doi = {10.1103/PhysRevD.110.L051301},
  url = {https://link.aps.org/doi/10.1103/PhysRevD.110.L051301}
}

@article{Sang2024,
  title = {},
  author = {Sang, Wen-Long and Feng, Feng and Jia, Yu},
  journal = {Phys. Rev. D},
  volume = {110},
  issue = {5},
  pages = {L051302},
  numpages = {6},
  year = {2024},
  month = {Sep},
  publisher = {American Physical Society},
  doi = {10.1103/PhysRevD.110.L051302},
  url = {https://link.aps.org/doi/10.1103/PhysRevD.110.L051302}
}

@article{Sang2025,
  author         = {Sang, Wen-Long and Feng, Feng and Jia, Yu},
  journal        = {},
  title          = {Mixed Electroweak-QCD Corrections to $H \to \gamma\gamma$},
  year           = {2025},
  eprint         = {2510.25516},
  archivePrefix  = {arXiv},
  primaryClass   = {hep-ph}
  }

@article{FeynArts,
title = {},
journal = {Computer Physics Communications},
volume = {140},
number = {3},
pages = {418-431},
year = {2001},
issn = {0010-4655},
doi = {https://doi.org/10.1016/S0010-4655(01)00290-9},
url = {https://www.sciencedirect.com/science/article/pii/S0010465501002909},
author = {Thomas Hahn}
}

@article{FeynCalc,
title = {},
journal = {Computer Physics Communications},
volume = {256},
pages = {107478},
year = {2020},
issn = {0010-4655},
doi = {https://doi.org/10.1016/j.cpc.2020.107478},
url = {https://www.sciencedirect.com/science/article/pii/S001046552030223X},
author = {Vladyslav Shtabovenko and Rolf Mertig and Frederik Orellana}
}

@article{FeynHelpers,
title = {},
journal = {Computer Physics Communications},
volume = {218},
pages = {48-65},
year = {2017},
issn = {0010-4655},
doi = {https://doi.org/10.1016/j.cpc.2017.04.014},
url = {https://www.sciencedirect.com/science/article/pii/S0010465517301285},
author = {Vladyslav Shtabovenko}
}

@article{PackageX,
title = {},
journal = {Computer Physics Communications},
volume = {197},
pages = {276-290},
year = {2015},
issn = {0010-4655},
doi = {https://doi.org/10.1016/j.cpc.2015.08.017},
url = {https://www.sciencedirect.com/science/article/pii/S0010465515003033},
author = {Hiren H. Patel}
}

@article{LoopTools,
title = {},
journal = {Computer Physics Communications},
volume = {118},
number = {2},
pages = {153-165},
year = {1999},
issn = {0010-4655},
doi = {https://doi.org/10.1016/S0010-4655(98)00173-8},
url = {https://www.sciencedirect.com/science/article/pii/S0010465598001738},
author = {T. Hahn and M. Pérez-Victoria}
}

@article{ATLASCMS2025,
    author = "Aad, Georges and others",
    journal = {},
    collaboration = "ATLAS, CMS",
    title = "{Highlights of the HL-LHC physics projections by ATLAS and CMS}",
    eprint = "2504.00672",
    archivePrefix = "arXiv",
    primaryClass = "hep-ex",
    reportNumber = "ATL-PHYS-PUB-2025-018, CMS-HIG-25-002",
    month = "4",
    year = "2025"
}

@article{FeynHiggs,
title = {},
journal = {Computer Physics Communications},
volume = {249},
pages = {107099},
year = {2020},
issn = {0010-4655},
doi = {https://doi.org/10.1016/j.cpc.2019.107099},
url = {https://www.sciencedirect.com/science/article/pii/S0010465519304059},
author = {H. Bahl and T. Hahn and S. Heinemeyer and W. Hollik and S. Paßehr and H. Rzehak and G. Weiglein}
}

@article{ATLASphph2023,
    author = "Aad, Georges and others",
    collaboration = "ATLAS",
    title = {},
    reportNumber = "CERN-EP-2022-094",
    doi = "10.1007/JHEP07(2023)088",
    journal = "Journal of High Energy Physics",
    volume = "07",
    pages = "088",
    year = "2023"
}

@techreport{ATLAS-CONF-2025-006,
      author = {},
      collaboration = "ATLAS",
      title         = "{Combined measurements of Higgs boson production and decay
                       at $\sqrt{s} =$ 13 TeV using up to 140 fb$^{-1}$ of data
                       collected by the ATLAS Experiment}",
      institution   = "CERN",
      reportNumber  = "ATLAS-CONF-2025-006",
      address       = "Geneva",
      year          = "2025",
      url           = "https://cds.cern.ch/record/2937634",
}

@article{Djouadi1998,
    author = "Djouadi, Abdelhak",
    title = {},
    eprint = "hep-ph/9806315",
    archivePrefix = "arXiv",
    reportNumber = "PM-98-06, GDR-S-06",
    doi = "10.1016/S0370-2693(98)00784-9",
    journal = "Phys. Lett. B",
    volume = "435",
    pages = "101--108",
    year = "1998"
}

@article{deFlorian2016,
  author       = "de Florian, Daniel and others",
  title        = {Handbook of LHC Higgs Cross Sections: 4. Deciphering the Nature of the Higgs Sector},
  journal      = {CERN Yellow Reports: Monographs},
  volume       = {2},
  pages        = {1--620},
  year         = {2017},
  doi          = {10.23731/CYRM-2017-002},
  eprint       = {1610.07922},
  archivePrefix= {arXiv},
  primaryClass = {hep-ph}
}

\end{document}